\documentclass[manuscript,screen]{acmart}
\AtBeginDocument{%
  \providecommand\BibTeX{{%
    \normalfont B\kern-0.5em{\scshape i\kern-0.25em b}\kern-0.8em\TeX}}}

\setcopyright{acmlicensed}
\copyrightyear{2024}
\acmYear{2024}

\makeatletter
\renewcommand\@formatdoi[1]{\ignorespaces}
\makeatother

\acmConference[CHI2024]{Make sure to enter the correct
  conference title from your rights confirmation email}{May 11-16, 2024}{Honolulu, HI, United States}
\acmBooktitle{CHI2024: With or Without Permission: Site-Specific Augmented Reality for Social Justice,
 May 11-16, 2024, Honolulu, HI, United States} 
\acmISBN{978-1-4503-XXXX-X/18/06}

\begin{document}

\title{\textbf{Augmented Voices: An Augmented Reality Experience Highlighting the Social Injustices of Gender-Based Violence in the Muslim South-Asian Diaspora}}

\author{Hamida Khatri}
\email{hamida.khatri@utdallas.edu}
\orcid{0000-0002-6596-8413}
\affiliation{%
  \institution{The University of Texas at Dallas}
  \city{Richardson}
  \state{Texas}
  \country{USA}
  \postcode{75080}
}


\begin{abstract}
  This paper delves into the distressing prevalence of gender-based violence (GBV) and its deep-seated psychological ramifications, particularly among Muslim South Asian women living in diasporic communities. Despite the gravity of GBV, these women often face formidable barriers in voicing their experiences and accessing support. "Augmented Voices" emerges as a technological beacon, harnessing the potential of augmented reality (AR) to bridge the digital and physical realms through mobile devices, enhancing the visibility of these often-silenced voices. With its technological motivation firmly anchored in the convergence of AR and real-world interactions, "Augmented Voices" offers a digital platform where storytelling acts as a catalyst, bringing to the fore the experiences shared by these women. By superimposing their narratives onto physical locations via Geographic Information System (GIS) Mapping, the application "augments their voices" in the diaspora, providing a conduit for expression and solidarity. This project, currently at its developmental stage, aspires to elevate the stories of GBV victims to a level where their struggles are not just heard but felt, forging a powerful connection between the user and the narrative. "Augmented Voices" is designed to transcend the limitations of conventional storytelling, creating an "augmented" reality where voices that are often muted by societal constraints can resonate powerfully. Through this intimate and immersive engagement, the project underscores the urgent imperative to confront GBV, catalyzing societal transformation and fostering robust support networks for those in the margins. The project is a pioneering example of how technology can become a formidable ally in the fight for social justice and the empowerment of the oppressed. Additionally, this paper delves into the nuanced AR workflow of "Augmented Voices," illustrating its relevance and contribution to the broader theme of site-specific AR for social justice.
\end{abstract}



\maketitle

\section{Introduction}
Gender-based violence (GBV) refers to any detrimental act committed against an individual's will, rooted in the societal roles assigned to different genders \cite{1}. This broad category encompasses various forms of violence such as domestic abuse, physical assault, rape, sexual harassment, female genital mutilation, enslavement, coerced prostitution, and involuntary or early marriages \cite{2,3}. While GBV can affect men, statistics show that globally, one in three women suffers from more intense sexual aggression, harsher physical violence, and greater domination by male counterparts \cite{1}. Studies indicate that women are up to eight times more likely to be victims of violence compared to men \cite{4}. The primary immediate impact on GBV victims is profound health and social repercussions, which subsequently lead to a variety of mental, emotional, and physical health challenges.

\section{Gender-Based Violence Within the Muslim South-Asian Diaspora}
GBV stands as a critical social issue within the United States (U.S.), raising national concern. Notably, within the Muslim South-Asian diaspora, cultural dynamics contribute to the underreporting of any violence against women due to structural inequities and societal disempowerment of victims. This underreporting leads to varied findings regarding the prevalence of GBV, with physical violence reported as the most common form (48\%), followed by emotional (38\%), economic (35\%), verbal (27\%), immigration-related (26\%), in-laws related (19\%), and sexual abuse (11\%), with higher prevalence rates among women for all violence types \cite{22}. Despite the South Asian community being one of the fastest-growing immigrant groups in the U.S., with an estimated population of 5.4 million, research on GBV victimization rates within this group has been limited \cite{23}.

 The nature of GBV within South Asian communities is influenced by cultural values such as patriarchy, collectivism, and male dominance, leading to unique manifestations of violence against women, including abuse by in-laws due to the common practice of living with extended family \cite{25}. Studies have shown significant emotional abuse by in-laws and highlighted the role of acculturation in influencing GBV victimization risks, with both low and high levels of acculturation having potential to increase GBV risks due to conflicting gender-role behaviors and challenges to traditional family structures \cite{25}.

 Furthermore, the collectivist nature of South Asian families affects marriage practices and the role of in-laws, often leading to abuse related to dowry demands and arranged marriages \cite{24,26}. For first generation immigrant women, the lack of a support network in the U.S. increases dependence on their husbands for economic, emotional, and immigration-related support, thereby exacerbating the risk of GBV victimization due to the power differential \cite{27}. Conversely, having social supports in the U.S. can serve as a protective factor by offering opportunities to disclose abuse and seek help, though informal support systems may sometimes discourage seeking external assistance \cite{28}. The complexity of these dynamics underscores the need for nuanced understanding and support strategies tailored to the unique experiences of GBV within the Muslim South-Asian diaspora.

\section{Storytelling as a Catalyst for Addressing Gender-Based Violence}
Traditionally, “disclosure” has been understood as the act of expressing one's innermost thoughts and emotions regarding distressing experiences either verbally or in writing \cite{31}. This concept has been particularly applied in the context of HIV, as well as to some extent in cases of GBV and other forms of victimization such as sexual abuse and dating violence \cite{32,33}. Research indicates that disclosing experiences of GBV can have positive outcomes, while the suppression of disclosure can act as a stressor \cite{33,34,35,36,37,38}.

 Traditional models of disclosure often depict the individual disclosing as solely responsible for conveying the event, but this overlooks the collaborative nature of the process, particularly in the context of GBV \cite{39}. Instead, a “discursive approach” to disclosure recognizes it as a shared construction of meaning between the individual disclosing and those supporting or listening \cite{39}. This approach emphasizes the role of language and social context in shaping what is disclosed and how it is understood. Furthermore, the act of disclosing GBV is not necessarily preceded by an awareness of the problem; rather, it may emerge through dialogue and mutual recognition of experiences \cite{42}. Additionally, disclosure and concealment of GBV are not mutually exclusive processes but may occur simultaneously and interactively \cite{40,41}.

 While existing research has examined factors influencing disclosure, there is a lack of studies specifically analyzing the process of GBV disclosure itself, particularly from the perspectives of both the victims and survivors \cite{42}. “Magical realism,” as a narrative strategy, is employed to represent unspeakable historical traumas, offering a tool termed "traumatic imagination," which facilitates the exploration and processing of trauma through empathetic storytelling \cite{44}. “Traumatic imagination” involves transforming traumatic memories into narratives of suffering through creative processes \cite{30}. Ayehsa et al. underscores the significance of storytelling in promoting the mental well-being of GBV victims and provides a literary perspective on how women's experiences of suffering are conveyed in traditional narratives \cite{30}. In contexts of conflict, storytelling becomes a crucial means of channeling trauma and fostering resilience. 

 Traditional storytelling, deeply rooted in oral traditions and poetry from the Indo-Persian or premodern era, serves as a resource for women who have experienced violence, enabling them to articulate their suffering and shield themselves from potential harm or stigma \cite{45}. These stories, often infused with magical realism, allow for the assimilation of memories and legacies, while also challenging dominant narratives and patriarchal structures \cite{46}.

 In the premodern era, storytelling was considered an art form aimed at transmitting knowledge and wisdom. Walter Benjamin, in his renowned 1936 essay "The Storyteller," emphasized the role of storytellers as conveyors of experience and insight \cite{47}. They passed down wisdom from generation to generation, encouraging their audience to maintain a sense of wonder and imagination. Utilizing their vast reservoir of memory, premodern storytellers recounted tales from ancient traditions, focusing not on practical solutions but on cultivating the ability to craft compelling narratives. This skill, according to Benjamin, allowed listeners to glimpse the richness of the world beyond mere scientific truths. 

 In Persianate India, storytelling served primarily as a tool for ethical development. Rather than being concerned with factual accuracy, these narratives focused on shaping individuals' inner selves and their relationships with others. The emphasis was on structuring human interiority and fostering moral growth. Magical realism, prevalent in Turkish literature, provides a platform for marginalized voices to be heard, evoking a sense of liminality that transcends boundaries and connects individual stories to broader human experiences \cite{46}. Through this narrative approach, experiences of trauma are transformed into narratives that offer meaning and healing for both storytellers and listeners. By integrating personal narratives with traditional stories, individuals and communities can engage in collective healing processes, drawing on family networks and community-based interventions that utilize storytelling in trauma therapy and recovery.

Storytelling vividly brings to light the deeply personal and often harrowing experiences of women who have endured any form of GBV. One particularly striking account from Kameela (pseudonym) a woman in Kabul captures the essence of such violence, not only in its physical form but also through emotional and verbal abuse. She poignantly states, "My husband doesn’t always beat me–but his attitude and the bad names he calls me is worse than if he stabs me with a sword. With this sense of fear, my whole body shakes, especially my hands and whatever is in my hand, falls down and breaks down. Then he gets so angry and beats me that why did I break glasses and household utensils. When I want to tell him that I didn’t break things intentionally, my tongue gets paralyzed and I can’t speak a word—it’s all the fear, I think it’s better to die once than dying every day" \cite{18}. This testimony reveals the severe impact of psychological torment and physical aggression. In another instance, Shazia (pseudonym), a 20-year-old woman from Pakistan, shares her ordeal, saying, "He grabbed me firmly, shoving me against the wall and unleashing a barrage of kicks and slaps. Then he picked up a metal pipe and started hitting me mercilessly" \cite{19}. Her words lay bare the brutal reality of the violence she faced.

The obstacles to obtaining justice and support for victims are further highlighted by the narratives of women discussing their lack of access to legal information. One woman, in a focused group discussion by the Asia Foundation, expressed, "There is no law for women like us. We are uneducated people and we don’t know anything about laws. We cannot read newspapers—maybe we could have read them if we had an education. There is no facility to go outside, so where would we go for information?" \cite{20}. This highlights a significant barrier to empowerment and protection against abuse. Adding to the discourse on gender inequality, another woman reflected on the societal attitudes that undermine women, "A man is a man; he considers himself everything. Men don’t respect women. It’s their prerogative whether they want to take us anywhere; otherwise, a woman cannot go anywhere by her own will. There is no difference between our animals and us." In a candid admission that illustrates the depth of gender bias, a man admitted, "We don’t seek the opinion of the woman, and even if she gives us some good advice, we ignore her, considering her less intelligent. After that, she feels she has no value in the household," \cite{20}.

Incorporating the voices of those who have experienced GBV deepens the narrative, bringing to light the societal norms and legal challenges faced by women in South Asian communities. Muskaan (pseudonym) shares her reluctance to disclose the violence within her marriage, reflecting on societal expectations: “I didn’t want to hurt [my family by disclosing the violence]. I thought it might change. Probably it won’t get to a point where I have to tell them. Qnukay (because) there’s norm like all the husbands beat their wives, not in my family, but . . .” \cite{21}. This highlights the normalization of domestic violence and the internal conflict faced by victims contemplating disclosure. The issue extends to the disturbing acceptance of honor killings in some South Asian societies, where female relatives can be murdered with societal and legal leniency under the guise of defending “family honor.” This grim reality is underscored by the fact that, "In our society, honor killing, a man can kill his wife and there’s laws to protect men,” revealing the legal protections afforded to perpetrators \cite{21}. Meena’s (pseudonym) experience of sexual violence on her wedding night further illustrates the normalization of abuse within marriage, as she believed such brutality was an expected part of marital relations: “First day, almost he raped me. I’m crying, illiterate, crying and, just he finished his work and I’m, I’m so, I felt so bad, but I think, this is supposed to happen” \cite{21}.

Preeti (pseudonym) recounts the exploitation she endured when an uncle, who moved into her family’s home under the pretext of immigration and establishing himself in the United States, abused her: “He was immigrating to the U.S. My father sponsored him to come. Until he could get on his feet and find a job and establish himself, he needed a place to stay...He had access to everything in our family. Our house was his house” \cite{21}. This account sheds light on the abuse of trust and power within familial settings. Kamelesh’s (pseudonym) story further illustrates the complex dynamics of abuse within the community, detailing the shift from affectionate attention to aggression by her uncle as she began to resist his advances: “My father is the eldest son...He’s everybody’s big guardian. I mean the community doesn’t even know how physically violent he can be...With my uncle, the abuse started when I was fifteen...It was only later that he began being aggressive with me, after I started to deny him access to me” \cite{21}. Her narrative highlights the contradiction between the community's perception of her family and her personal experiences of abuse.

These narratives shed light on the complex layers of various forms of GBV, from physical abuse to the psychological aftermath and the systemic barriers that prevent women from seeking help. They underscore the critical need for societal change to ensure the safety and dignity of all women.

Imagine a world where the complex layers of a victim's story, enriched with tangible evidence, are accessible through the simple touch of a mobile app. Here, the potent capabilities of AR give rise to a newfound empowerment, amplifying the voices of women striving for justice. Imagine each narrative, once muffled by the din of neglect, now resounding with clarity and strength, guiding us toward a reality where justice is not an abstract ideal but an attainable outcome. This is the driving force behind "Augmented Voices."

\section{"Augmented Voices" - Combining Augmented Reality with Social Justice Storytelling}
"Augmented Voices" is a pioneering mobile augmented reality (AR) application solution that intricately blends the historical and present-day expereinces of social injustice into the Muslim South Asian GBV victim’s real-world surroundings. This app embarks users on an engaging exploration of their environment, transforming specific locales into catalysts for unveiling stories, immersive visuals, and poignant soundscapes that collectively portray the struggles and resilience of marginalized South Asian Muslim women across various geographies. Though still in the concept development stage, the content of "Augmented Voices" will carefully be selected from a wide array of sources, encompassing archival documents, interviews, direct personal narratives submitted via the Mobile App, and seminal texts on social justice. This eclectic collection of stories will deliver a rich and nuanced understanding of injustice and endurance, providing users with deep insights into the lives of those who have stood up to systemic oppression and violence from a Muslim South Asian perspective. By presenting stories from historical episodes of discrimination to modern-day accounts of GBV, the app aims to bridge the temporal divide, shedding light on the persistent nature of these conflicts while celebrating the courage and tenacity of affected individuals and groups. 

"Augmented Voices" will leverage the cutting-edge AR technology to generate dynamic, context-aware visualizations that respond to the user's specific location and movements, significantly enriching the narrative immersion. This interactive dimension will not only amplify the narrative's impact but also motivate users to physically engage with their surroundings, thus enhancing their connection to the stories shared. The application's sound design will be integral in crafting this immersive environment, with a selection of audio pieces that perfectly accentuate the visual narrative, inviting users into the emotional and psychological realms of the stories.

As an innovative educational and advocacy instrument, "Augmented Voices" hopes to play a vital role in raising awareness of the battles against social injustice with respect to GBV of the Muslim South Asian groups. It hopes to cultivate empathy and unity among users, prompting them to contemplate their contribution to uplifting and echoing the voices of marginalized individuals within their communities.

\section{Technological Motivation}
AR represents a transformative technology at the intersection of the physical and digital worlds, adding a meaningful layer of content that enriches real-world environments. In the context of addressing GBV against Muslim South Asian women—often marginalized and voiceless—AR offers a powerful medium for engagement and advocacy. "Augmented Voices" will utilize Geographic Information System (GIS) mapping to anchor storytelling in specific geographical contexts, thereby creating a deep, situative experience for users.

The core technological motivation behind integrating AR with storytelling in this social issue context lies in its ability to combine the tangible physical world with enriched, contextually relevant digital information. This blend facilitates meaningful experiences situated within the physical environments where these narratives unfold or are rooted. For victims of GBV, particularly Muslim South Asian women, such digital enhancement is not merely narrative but serves as a potent tool for awareness and understanding.

According to Hope Schroeder, Rob Tokanel, Kyle Qian, and Khoi Le (2023), modern AR systems possess the capability to recognize locations ("where things are") but often lack the depth to understand the significance ("what they are" and "why they matter") of these places \cite{48}. This insight underlines the challenge and the potential of AR: to provide an "overlay on reality" where the physical presence of a place is augmented with historical, cultural, or personal significance through digital media.

For "Augmented Voices," visual markers will be used to trigger location-based AR experiences. This will ensure that interactions are not just immersive but are integrally tied to the real-world locations that hold significance in the narratives of GBV. By leveraging AR's capability to utilize 3D space for site-specific storytelling, the app not only hopes to tell a story, but it allows users to step "into the story," enhancing engagement through interactive and immersive elements.

Looking ahead, the development of site-specific AR for "Augmented Voices" is in the planning phase and will progress over the coming months. The aim is to employ AR not just as a narrative tool but as a medium for "traumatic imagination" where users can confront and comprehend the realities of GBV in a visceral, impactful manner. This approach intends to transform how stories of social injustice and resilience are told and experienced, making AR a critical tool in the fight against GBV and in amplifying the voices of those who have been silenced.

\section{Geographic Information System (GIS) Mapping}
The GIS Mapping component of "Augmented Voices" plays a pivotal role in providing a spatial dimension to the storytelling and engagement process. This integration allows users to experience a deeper connection with the narratives by situating them within specific geographic contexts. The workflow or systems architecture incorporating GIS mapping into the "Augmented Voices" app involves several key steps, designed to offer an intuitive and immersive user experience.

\begin{figure}
    \centering
    \includegraphics[width=0.75\linewidth]{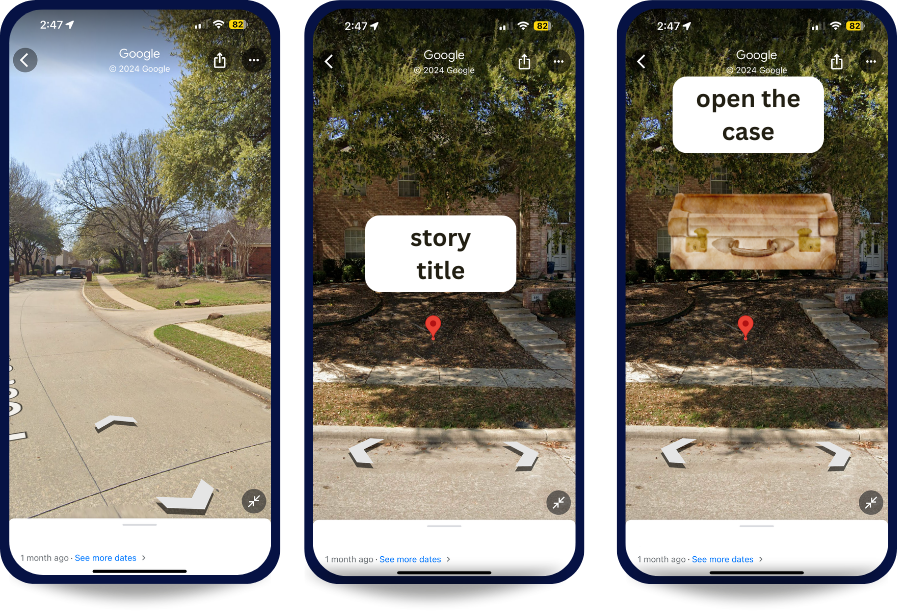}
    \caption{AR Story: represents the initial point where the story is marked and identified within the AR environment.}
    \label{fig:1}
\end{figure}
\begin{figure}
    \centering
    \includegraphics[width=0.75\linewidth]{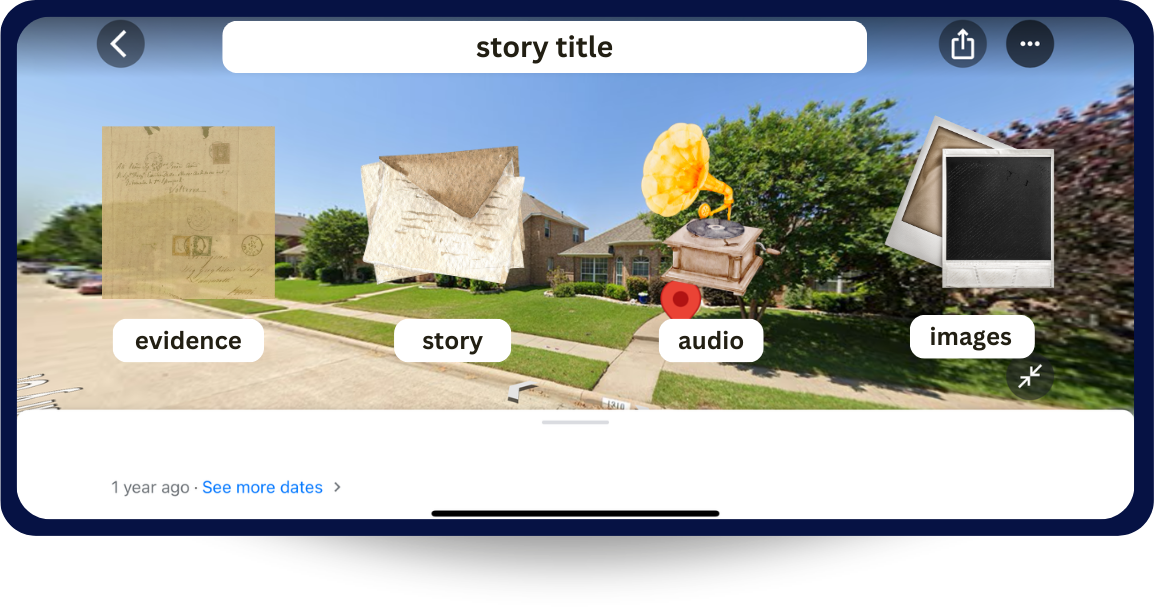}
    \caption{Interactive AR Case Files for the Story: indicates the options for users to explore different aspects of the story, such as evidence and audio recordings, in a more interactive and detailed manner.}
    \label{fig:2}
\end{figure}

\begin{figure}
    \centering
    \includegraphics[width=0.75\linewidth]{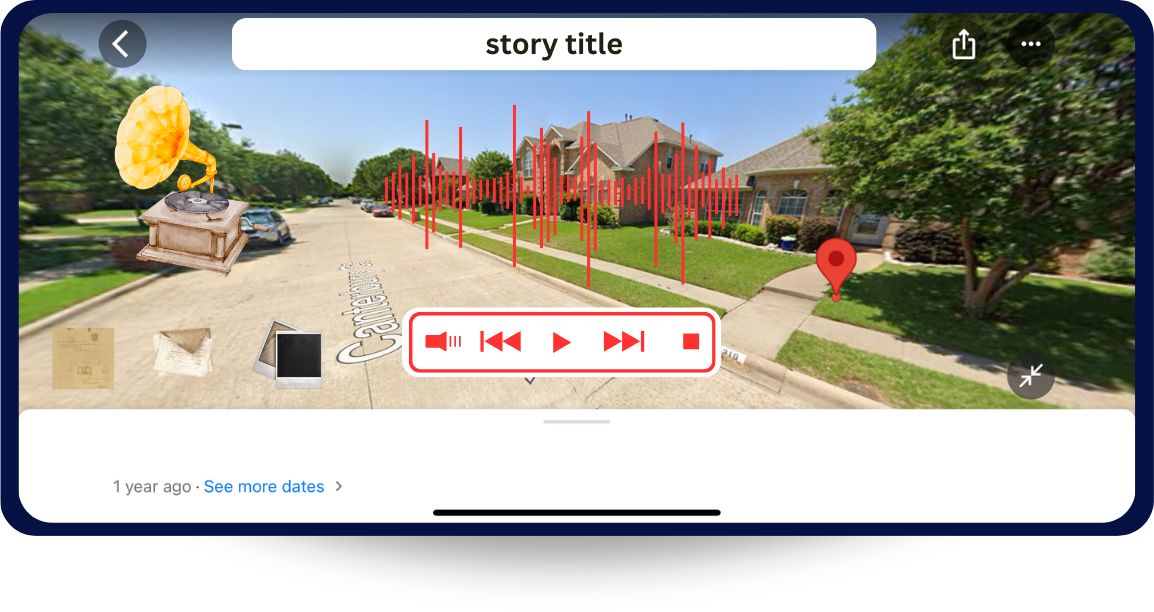}
    \caption{Expanded Audio Soundscape: audio feature within the app providing a more immersive experience of the story through sound.}
    \label{fig:3}
\end{figure}

 \textbf{
\subsection{User Onboarding and Sign-up Process}
}

1. \textbf{User Onboarding:} Upon downloading the "Augmented Voices" app, users are greeted with an onboarding process that introduces the app's mission, features, and how to navigate through its content. This stage is crucial for setting the context and preparing users for the interactive experience.

2. \textbf{Sign-up:} Users are prompted to create an account to personalize their experience. The sign-up process involves entering basic information and preferences, which could include interests in specific historical periods, regions, or types of social injustices. This data helps in curating content that aligns with the user's interests.

 \textbf{
\subsection{Content Discovery and GIS Mapping}
}

1. \textbf{Discovery of Narratives:} After the initial setup, users are directed to the app's main interface, where they can explore various narratives. The app employs GIS mapping to display stories relevant to the user's current location or locations of interest they select on the map.

2. \textbf{Location-based Triggers:} As users move within their environment, the app uses their geographical location to trigger site-specific stories. These triggers are based on proximity to locations that are significant to the narratives within the app. When a user approaches a trigger point, the app notifies them and presents the opportunity to engage with a story associated with that location.

 \textbf{
\subsection{Interactive Experience}
}
1. \textbf{Immersive Storytelling:} Once a narrative is triggered, users are presented with immersive visualizations, detailed historical accounts, and evocative soundscapes that bring the story to life. This multidimensional approach, combining text, audio, and visual elements, enriches the user's understanding and emotional connection to the narrative.

2. \textbf{Real-time Interaction:} The GIS mapping component is designed to offer dynamic, context-sensitive visualizations that adapt to the user's location and movements. This real-time interaction encourages users to explore their physical surroundings, enhancing engagement with the content.

 \textbf{
\subsection{User Engagement and Community Building}
}

1. \textbf{Community Engagement:} Users can share their experiences and insights gained from the narratives with the "Augmented Voices" community. The app may feature discussion forums, social media integration, or options to share stories and locations with others, fostering a sense of solidarity and collective learning.

2. \textbf{Feedback and Personalization:} The app incorporates mechanisms for users to provide feedback on the stories and their overall experience. This feedback is used to refine and personalize future content, ensuring the app remains relevant and impactful for its user base.

 \textbf{
\subsection{User Story Submission}
}
 Adding a user-generated content feature to the "Augmented Voices" app significantly enhances its interactive and community-building aspects. This functionality allows users, who are constrained to be Muslim South-Asian demographic, to contribute their own stories or narratives related to social injustices and GBV, which then become part of the app's extensive storyboard. Here’s how this process integrates into the system's architecture and user workflow:

 \hspace{1cm}

1. \textbf{Narrative Contribution:} Users can submit their own stories directly through the app. This process includes filling out a submission form where they can detail their narrative, attach relevant media (photos, videos, audio clips), and specify the geographic location associated with their story.

2. \textbf{Moderation and Approval:} To maintain the integrity and sensitivity of the content, each submission goes through a moderation process. The app's team reviews the stories to ensure they align with the app's mission and content guidelines before approval.

3. \textbf{Geolocation Tagging:} As part of the submission process, users tag their stories with specific geographic locations. This tagging is crucial for integrating the narrative into the GIS mapping system, allowing the story to be triggered by future users who visit or express interest in that location.

 \textbf{
\subsection{Integration into GIS Mapping}
}

1. \textbf{Story Mapping:} Once approved, the user-generated narratives are added to the app's geolocated map. These contributions are visually represented on the map, making it easy for users to discover new, community-added stories alongside the existing curated narratives.

2. \textbf{Dynamic Content Discovery:} The inclusion of user-generated stories enriches the app's content library, offering a broader spectrum of experiences and perspectives. Users exploring the map can trigger these narratives in the same manner as the curated content, based on their physical location or selected areas of interest.

 \textbf{
\subsection{Engagement and Community Feedback}
}

1. \textbf{Interactive Engagement:} Users can interact with these community-contributed narratives by reading, listening to, or viewing them. Additionally, they have the option to leave comments, share reactions, or even share the narratives on social media, fostering a supportive community atmosphere.

2. \textbf{Community Storyboard:} All narratives, including user submissions, are displayed on a collective storyboard within the app. This storyboard serves as a central hub for discovering the diverse range of experiences and stories related to social injustices and gender-based violence, mapped across different geographic locations.

 \textbf{
\subsection{Personal and Collective Narratives}
}

1. \textbf{Empowering Voices:} By enabling users to add their stories, "Augmented Voices" not only amplifies individual voices but also strengthens the collective narrative around social injustices and resilience. This participatory approach empowers users, making them active contributors to the app's mission of raising awareness and fostering empathy.

 2. \textbf{Continuous Growth:} The feature ensures the app's content remains dynamic and evolving, reflecting the ongoing and diverse nature of social injustices globally. It encourages users to regularly engage with the app, exploring new stories and contributing their narratives, thereby continuously enriching the community's collective knowledge and empathy.

 \hspace{1cm}
 
 Integrating user-generated content through GIS mapping into "Augmented Voices" leverages the power of community storytelling, making the exploration of social injustices deeply personal yet universally accessible. This functionality will not only diversify the content available on the app but also foster a stronger, more engaged community committed to highlighting and combating GBV and social injustice worldwide.

\hspace{1cm}

\section{Relevance to Social Justice and Augmented Reality}
The "Augmented Voices" initiative represents a pivotal intersection of technological innovation and social justice, harnessing AR's immersive potential to spotlight the experiences of marginalized groups. By seamlessly integrating AR with stories of adversity, resilience, and victory, the project transcends mere digital innovation to become a crucial instrument for fostering empathy and elevating awareness. It transforms impersonal data and historical accounts into vivid, interactive experiences that connect users emotionally with the narratives, effectively narrowing the divide between them and the firsthand experiences of social injustice.

 Crucially, the initiative offers a unique opportunity to engage with stories and viewpoints often sidelined in mainstream discussions. Through AR, users transition from passive information receivers to active story participants, immersing them in the harsh realities of discrimination, inequality, and violence still prevalent today. This profound engagement serves as an empathy catalyst, placing users within the very context of injustice, thereby promoting a more intimate understanding and connection with those who face these adversities.

 Moreover, "Augmented Voices" underscores AR's potential as an educational and activist tool. It exemplifies how advanced technology can facilitate social good, creating an interactive experience that educates and motivates users toward societal discourse and action. By rendering concepts of justice and equality more tangible and comprehensible, the project prompts users to introspect about their societal roles and how they might contribute to advocating for transformative change.

 At its core, "Augmented Voices" demonstrates the potential of technological progress, especially AR, to advance social justice. It introduces an innovative approach to activism and education, leveraging immersive storytelling's power to kindle empathy, awareness, and proactive engagement. The project not only sheds light on the plights of marginalized South Asian Muslims but also emphasizes technology's role as a catalyst for positive societal change towards a more equitable and just world.

\section{ Future Considerations}
Moving forward with the project, a multi-faceted approach is essential to ensure the well-being of users, especially considering the sensitive nature of GBV narratives. A content warning system could be introduced, giving users prior notice about the emotional impact of AR experiences and allowing them to engage with the content at their comfort level. The integration of immediate mental health resources, such as crisis hotlines and counseling services, directly within the app would provide crucial support for GBV victims and survivors. Moderation policies could also be put in place for story submissions to prevent retraumatization and protect the privacy of individuals, especially in ongoing abuse scenarios. This could include anonymous participation options to encourage sharing without fear of identification or repercussion. Incorporating local resources, like domestic violence hotlines and organizational contacts, tailored to the user's location, will provide practical support as well. Educational outreach initiatives could offer vital information on recognizing abuse and assisting others, thereby extending the app's influence beyond storytelling into actionable awareness. Feedback mechanisms are critical to the iterative improvement of the app, ensuring that the content remains sensitive to the victim and survivor needs while supporting their healing process. Ethical storytelling guidelines could be established to protect the dignity of the narratives shared. Lastly, conducting long-term impact studies could help in understanding the effect of such AR experiences on users and guide the responsible evolution of the app as a tool for awareness, education, and advocacy.

\section{Conclusion}
In conclusion, the "Augmented Voices" project represents a pioneering fusion of AR, technology and social justice advocacy, offering a new paradigm for engaging with and understanding the narratives of marginalized communities especially Muslim South Asian women in the diaspora. Through its innovative use of AR to create immersive, geolocated storytelling experiences, this project transcends traditional educational methods, providing users with a deeply personal and empathetic insight into the realities of social injustice and inequality. By enabling users to not only consume but also contribute to the narrative landscape through the addition of their own stories, "Augmented Voices" democratizes the storytelling process, ensuring a diverse and authentic representation of experiences. This participatory element enriches the project's narrative tapestry and fosters a community of engaged and informed advocates for social change.

The significance of "Augmented Voices" lies not only in its technological achievements but also in its commitment to leveraging these advancements for the greater good. It challenges users to move beyond passive observation, inviting them to become active participants in a global dialogue on social justice. This project underscores the potential of AR as a powerful tool for education, activism, and empathy-building, encouraging users to reflect on their roles within larger social structures and to take action towards fostering a more equitable world.

As we look toward the future, the "Augmented Voices" project serves as a beacon of hope and a model for how technology can be harnessed to amplify marginalized voices, bridge cultural and geographical divides, and inspire collective action. It is a testament to the transformative power of combining innovation with compassion, highlighting the critical role of technology in advancing the cause of social justice and equity. Through "Augmented Voices," we are reminded of the enduring impact of storytelling as a means to connect, educate, and mobilize communities around the globe in the pursuit of a more just and inclusive society.

\bibliographystyle{ACM-Reference-Format}
\bibliography{references.bib}

\end{document}